\documentclass[twocolumn,showpacs,floatfix,superscriptaddress,amsmath,amssymb,prb]{revtex4}
\usepackage{bbm}
\usepackage{amsmath}
\usepackage{mathrsfs}
\usepackage{txfonts}
\usepackage{amssymb}
\usepackage{graphicx}
\usepackage{hyperref}
\usepackage{ulem}
 \usepackage{overpic}
 \usepackage{psfrag}
 \newcommand{\be}{\begin{equation}}
\newcommand{\ee}{\end{equation}}

\begin{document}
\title{Ground-state fidelity and tensor network states for quantum spin tubes}
\author{Ai-Min Chen, Qian-Qian Shi, Jin-Hua Liu and Huan-Qiang Zhou }

\affiliation{Centre for Modern Physics and Department of Physics,
Chongqing University, Chongqing 400044, The People's Republic of
China}

\date{\today}

\begin{abstract}
An efficient algorithm is developed for quantum spin tubes in the
context of the tensor network representations. It allows to
efficiently compute the ground-state fidelity per lattice site,
which in turn enables us to identify quantum critical points, at
which quantum spin tubes undergo  quantum phase transitions. As an
illustration, we investigate the isosceles spin 1/2
antiferromagnetic three-leg Heisenberg tube. Our simulation results
suggest that two Kosterlitz-Thouless transitions occur as the degree
of asymmetry of the rung interaction is tuned, thus offering an
alternative route towards a resolution to the conflicting results on
this issue arising from the density matrix renormalization group.
\end{abstract}

\maketitle

\section{Introduction}

Recently, significant progress has been made in developing efficient
numerical algorithms in the context of the tensor network (TN)
representations~\cite{mps,vidal1,tn1,tn2,tn3,tn5,PEPS,iPEPS,tn4,tn6},
as a consequence of our deeper understanding of the role of quantum
entanglement and fidelity in characterizing critical phenomena in
quantum lattice systems. Actually, these developments provide
significant insights into the working principle of the density
matrix renormalization group (DMRG)~\cite{DMRG}, which in turn has
led to some novel algorithms based on the TN representations. Among
them are the matrix product states (MPS)~\cite{mps} in one spatial
dimension and the projected entangled-pair states (PEPS)~\cite{PEPS}
in two or higher spatial dimensions. Especially, an infinite MPS
(iMPS) algorithm~\cite{vidal1} and an infinite PEPS (iPEPS)
algorithm~\cite{iPEPS} have been developed to compute the TN
representations of ground-state wave functions for translationally
invariant infinite-size quantum systems in one and two or higher
spatial dimensions, respectively. A remarkable feature of these TN
algorithms is that they offer an efficient way to compute the
ground-state fidelity per lattice site~\cite{zhou,zov}. Indeed,
quantum fidelity allows to capture the ground-state phase diagrams
for various quantum lattice systems in condensed
matter~\cite{zhou,zov,fidelity0,fidelity1,fidelity2,fidelity3}.

There are a class of quantum lattice systems lying between one and
two spatial dimensions, namely, quantum spin ladders and tubes. Both
experiments and theories~\cite{Lieb} have confirmed that the spin
1/2 Heisenberg ladders are gapful for an even number of legs and
gapless for an odd number of legs. In fact, the ground state of an
odd-leg Heisenberg spin ladder is a gapless spin-liquid or a
Tomonaga-Luttinger liquid~\cite{Luttinger}. Once the periodic
boundary conditions are applied in the rung direction, spin ladders
become spin tubes. Some exotic phenomena occur due to the geometric
frustration for an odd-leg spin tube, which has been realized in
some materials, e.g., for three-leg~\cite{three} and
nine-leg~\cite{nine} spin tubes. However, little attention has been
paid to these systems in the context of the TN representations.
Given the importance of both the spin ladders and spin tubes in the
theory of strongly correlated systems, it is desirable to develop TN
algorithms as an alternative means to investigate their fascinating
physics.

This paper is a part of our efforts to develop an efficient
numerical algorithm for quantum spin tubes in the context of the TN
representations, which offers an efficient way to compute the
ground-state fidelity per lattice site. Thus, we are able to locate
quantum criticalities for quantum spin tubes. As an illustration, we
investigate the isosceles spin 1/2 antiferromagnetic three-leg
Heisenberg tube. The simulation results suggest that two
Kosterlitz-Thouless (KT) transitions occur as the degree of
asymmetry of the rung interaction is tuned.

\section{tensor network states for quantum spin tubes}
%%%%%%%%%%%%%%%%%%%%%%%%%Fig1%%%%%%%%%%%%%%%%%%%%%%%%%%%%%%%%%%%
\begin{figure}
\centering
\includegraphics[angle=90,width=3in]{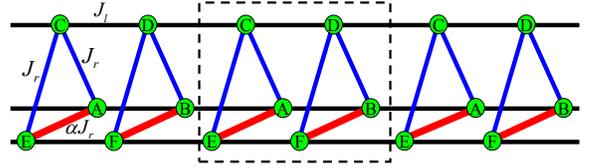}
\caption{(Color online) A three-leg spin tube with  exchange
interaction constants $J_l$ and $J_r$ along the leg and rung
directions, respectively. Here, an asymmetry parameter $\alpha$ is
introduced as a tunable control parameter. One choice of the unit
cell is highlighted in a dash-line box.}\label{model}
\end{figure}
%%%%%%%%%%%%%%%%%%%%%%%%%%%%%%%%%%%%%%%%%%%%%%%%%%%%%%%%%%%%%%%%%%

\begin{figure}
\centering
\includegraphics[angle=90,width=3in]{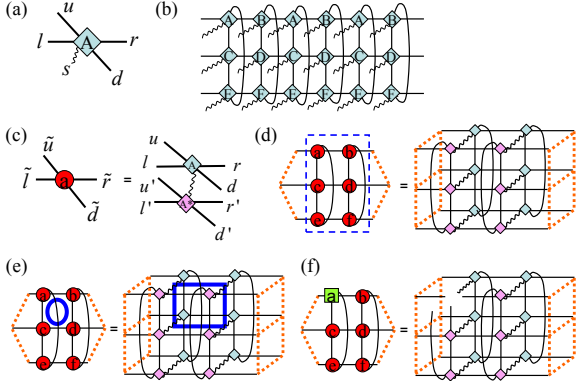}
\caption{(Color online) A TN representation for an infinite-size
three-leg spin tube. (a) A five-index tensor is attached to each
site, with $l$, $r$, $u$ and $d$ denoting the inner indices and $s$
the physical index. (b) A TN representation for a wave function,
which is translationally invariant under two-site shifts along the
leg direction. (c) An eight-index double tensor $a_{\tilde{l}
\tilde{r}\tilde{u}\tilde{d}}$ is formed, by contracting the physical
indices, from the five-index tensor $A_{lrud}^s$ and its complex
conjugate $A_{l'r'u'd'}^{s \ast}$, with $\tilde{l} \equiv (l,l'),
\tilde{r} \equiv (r,r'), \tilde{u} \equiv (u,u')$, and $\tilde{d}
\equiv (d,d')$. (d) The norm $\langle\psi|\psi\rangle$ of a given
wave function in the TN representation. The environment tensors are
nothing but the left and right dominant eigenvectors of the transfer
matrix  consisting of six double tensors. Here, the transfer matrix
is highlighted in the dash-line box. (e) The expectation value of a
physical observation, such as the four-site Hamiltonian density. (f)
The gradient $\partial\langle\psi |\psi\rangle/\partial A_{lrud}^{s
\ast}$ is equal to the norm with a tensor $A_{lrud}^{s \ast}$
absent. }\label{tensor}
\end{figure}

Assume that the model Hamiltonian $H$ describes the nearest-neighbor
interaction and is translationally invariant along the leg
direction. For a 3-leg spin tube, there are six sites $A$, $B$, $C$,
$D$, $E$ and $F$ in each unit cell, as shown in Fig.~\ref{model}.
Such a choice of the unit cell is necessary to accommodate a ground
state with spontaneously broken translational symmetry, as it does
occur in spin tubes. Indeed, there are two different choices of the
unit cell: one is $ABCDEF$, as highlighted in Fig.~\ref{model}, the
other is $BADCFE$. Any wave function admits a TN representation,
which follows from attaching to each site a five-index tensor, as
visualized in Fig.~\ref{tensor}(a) for a five-index tensor
$A_{lrud}^{s}$. Here, $s$ $(s=1,2,\cdots,\mathbbm{d})$ is a physical
index from a $\mathbbm{d}$-dimensional local Hilbert space. $l$,
$r$, $u$ and $d$ denote the bond indices from the $\chi$-dimensional
auxiliary spaces, with $\chi$ being the truncation dimension, as
depicted in Fig.~\ref{tensor}(b). In order to measure a physical
observable, we define the eight-index double tensors, which are
formed, by contracting the physical indices, from the five-index
tensors and their complex conjugates,  as shown in
Fig.~\ref{tensor}(c). Fig.~\ref{tensor}(d) shows how to compute the
norm $\langle\psi|\psi\rangle$ for a given wave function in the TN
representation. Here, a key notion is the transfer matrix, which is
highlighted in the dash-line box. The  dominant left and right
eigenvectors of the transfer matrix are nothing but the environment
tensors, in the sense that it accommodates the effect of the
remaining part of the infinite-size spin tube system, once the
system is partitioned into two parts, with one of them being the
transfer matrix. Notice that the norm follows from the dominant
eigenvalue $\lambda_1$ of the transfer matrix.
%%%%%%%%%%%%%%%%%%%%%%%%%Fig2%%%%%%%%%%%%%%%%%%%%%%%%%%%%%%%%%%%

%%%%%%%%%%%%%%%%%%%%%%%%%%%%%%%%%%%%%%%%%%%%%%%%%%%%%%%%%%%%%%%%%%

\section{Updating procedure for the tensor Network}
We need to update all the tensors  in the unit cell during the
iteration to generate the ground-state wave function for a given
Hamiltonian. Here, we focus on how to update the tensor
$A_{lrud}^{s}$ to explain the procedure. In fact, other tensors are
updated in exactly the same way.

For a given quantum state $|\psi\rangle$, the energy per site, $e$,
is written as
\begin{equation}
E= \lim_{L \rightarrow \infty} \frac {1}{L} \; \frac{ \langle\psi |
H |\psi\rangle}{ \langle\psi | \psi\rangle}.
\end{equation}
Here, $L$ is the total number of the lattice site. In the TN
representation, the energy per site, $e$, is a functional of the TN
tensors, which may be computed efficiently, as shown in
Fig.~\ref{tensor}(e).

\begin{figure}
\centering
\includegraphics[angle=90,width=3in]{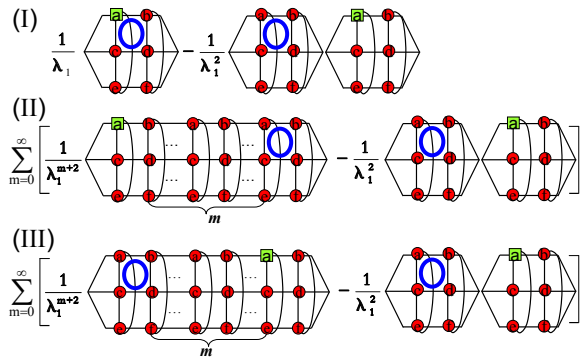}
\caption{(Color online) The TN representation for the energy
gradient $G^A$ with respect to the tensor $A^{\ast}$. A small circle
represents a double tensor attached to each site, a square indicates
the absence of the tensor $A^{\ast}$, and a large circle is the
Hamiltonian density acting on a plaquette in the unit cell. Here,
$\lambda_1$ denotes the dominant eigenvalue of the transfer matrix.
The TN for the energy gradient consists of three parts: (I) the hole
and the Hamiltonian density posit on the same cell; (II) and (III)
the Hamiltonian cell posits on the left and right hand sides of the
hole cell, respectively. Here, $m$ is the cell number between the
hole cell and the Hamiltonian cell. Notice that only the
contribution from the plaquette Hamiltonian density $h_{ABCD}$ is
shown.  }\label{grad}
\end{figure}

Now we turn to the computation of the gradient of the energy
functional with respect to $A^{\ast}$:
\begin{equation}\label{gradient}
G^A=\lim_{L\rightarrow \infty} \frac{1}{L} [\frac{1}{\langle\psi
|\psi\rangle}\frac{\partial\langle\psi| H |\psi\rangle}{\partial
A^{\ast}}-\frac{\langle\psi | H |\psi\rangle}{\langle\psi
|\psi\rangle^2} \frac{\partial\langle\psi|\psi\rangle}{\partial
A^{\ast}}],
\end{equation}
where $A^{\ast}$ is the complex conjugate of $A$. Here and
hereafter, we have omitted the tensor indices for clarity.  In order
to compute the energy gradient efficiently, we take advantage of the
translational invariance of the Hamiltonian. Since there are two
different choices of the unit cell, the contribution to the energy
gradient involves six Hamiltonian densities acting on different
plaquettes, i.e., $h_{ABCD}$, $h_{BADC}$, $h_{CDEF}$, $h_{DCFE}$,
$h_{EFAB}$ and $h_{FEBA}$.  Then the energy gradient becomes
\begin{equation}\label{gradient1}
G^A=\sum_m G^A_m,
\end{equation}
with
\begin{equation}\label{gradient2}
G^A_m=\frac{1}{\langle\psi |\psi\rangle}\frac{\partial\langle\psi|
h_m |\psi\rangle}{\partial A^{\ast}}-\frac{ \langle\psi | h_m
|\psi\rangle}{\langle\psi|\psi \rangle^2}
\frac{\partial\langle\psi|\psi \rangle}{\partial A^{\ast}}.
\end{equation}
Here, $h_m \in \{ h_{ABCD}, h_{BADC}, h_{CDEF}, h_{DCFE}, h_{EFAB},
h_{FEBA} \}$. As an example, we take the Hamiltonian density
$h_{ABCD}$ to explain how to compute the gradient. As shown in
Fig.~\ref{grad},  there is a TN representation for the gradient
$G^A_{ABCD}$, in which a hole cell is the cell with the tensor
$A^{\ast}$ absent, and a Hamiltonian cell is the cell with the
Hamiltonian density $h_{ABCD}$ sandwiched by attaching to the
physical indices between four five-index tensors and their complex
conjugates. The expectation value of the Hamiltonian density
$h_{ABCD}$ and the gradient
$\partial\langle\psi|\psi\rangle/\partial A^{\ast}$ are shown in
Fig.~\ref{tensor}(e) and (f), respectively. Notice that the gradient
$\partial\langle\psi |\psi\rangle/\partial A^{\ast}$ is nothing but
the norm with a tensor $A^{\ast}$ removed. In addition,
$\partial\langle\psi| h_m |\psi\rangle/\partial A^{\ast}$is nothing
but the expectation value of $h_m$ with the tensor $A^{\ast}$
removed. Depending on the relative locations of the hole and the
Hamiltonian cells, the TN representation for the energy gradient
consists of three parts: (I) the hole and the Hamiltonian density
posit on the same cell; (II) the Hamiltonian cell posits on the left
hand side of the hole cell; and (III) the Hamiltonian cell posits on
the right hand side of the hole cell. It should be emphasized that
the gradient decays very fast when the Hamiltonian cell falls apart
from the hole cell. That is, only a few Hamiltonian cells need to be
taken into account to meet a preset precision. Similarly, one may
compute the contribution to the gradient $G_A$ from other
Hamiltonian densities.

%%%%%%%%%%%%%%%%%%%%%%%%%Fig3%%%%%%%%%%%%%%%%%%%%%%%%%%%%%%%%%%%

%%%%%%%%%%%%%%%%%%%%%%%%%%%%%%%%%%%%%%%%%%%%%%%%%%%%%%%%%%%%%%%%%%

Once one knows the energy gradient,  the real and imaginary parts,
$A_{re}$ and $A_{im}$, of the tensor $A$ may be updated as follows:
\addtocounter{equation}{1}
\begin{align}\label{update}
\tilde{A}_{re}&=A_{re}-\delta~
 \frac{G^A_{re}}{|G^A_{re}|_{\rm{max}}} ~ , \tag{\theequation a}\\
\tilde{A}_{im}&=A_{im}-\delta~
\frac{G^A_{im}}{|G^A_{im}|_{\rm{max}}} ~ , \tag{\theequation
b}\label{update2}
\end{align}
where $G_{re}^A$ and $G_{im}^A$ are, respectively, the real and
imaginary parts of the five-index tensors $G^A$, $\delta$ is the
step size for an update. During the implementation, it is tuned to
be decreasing with the number of the iteration steps. Notice that
both the sign and magnitude of the energy gradients $G^A_{re}$ and
$G^A_{im}$ are exploited. As such, the tensor $A$ is updated to
$\tilde{A}=\tilde{A}_{re}+ i \tilde{A}_{im}$. Actually, all the six
tensors in the unit cell are updated simultaneously. Repeating the
procedure until the ground-state energy per site converges, an
approximate ground-state wave function is generated in the TN
representation.

\section{Model}

We consider the antiferromagnetic spin 1/2 three-leg Heisenberg
tube, as depicted in Fig.~\ref{model}, on an infinite-size lattice.
The Hamiltonian takes the form

\addtocounter{equation}{1}
\begin{align}
\hat{H}&=\hat{H}_{\rm{leg}}+\hat{H}_{\rm{rung}}, \label{ham2} \tag{\theequation a}\\
  \hat{H}_{\rm{leg}}&=J_{l}~\sum_{i,j}~\textbf{S}_{i,j}~ \cdot~
  \textbf{S}_{i+1,j}, \tag{\theequation b}\\
  \hat{H}_{\rm{rung}}&= J_{r}~ \sum_{i}~(~\textbf{S}_{i,1}~
  \cdot~ \textbf{S}_{i,2}~+~
  \textbf{S}_{i,2}~ \cdot~ \textbf{S}_{i,3}+~ \alpha ~\textbf{S}_{i,3}~ \cdot~
  \textbf{S}_{i,1}~), \tag{\theequation c}
\end{align}
where $\textbf{S}_{i,j}$ is the spin-$1/2$ Pauli operators at leg
$i$ and rung $j$ $(j=1,2,3)$, $J_l$ is the neighboring exchange
interaction along the legs and $J_r$ stands for the rung
interaction. For the antiferromagnetic coupling, the model is
frustrated along the rung direction. The asymmetric parameter
$\alpha$ controls the strength of the frustration. If the asymmetric
parameter $\alpha$ is varied, the model exhibits two limiting cases:
for $\alpha=0$, it corresponds to the three-leg ladder; for
$\alpha\rightarrow\infty$, it becomes a decoupled system consisting
of a single chain and a two-leg ladder. Both of them are gapless
Tomonaga-Luttinger liquids, with the central charge $c=1$, and
possess the $SU(2)$ symmetry. When $\alpha=1$, the system is
spontaneously dimerized, arising from the geometric frustration. All
the spin excitations are gapful, as follows from the
Lieb-Schultz-Mattis theorem~\cite{LSM}:  two degenerate ground
states arise from the broken translational symmetry. Therefore, one
may expect that at least one phase transition point occurs if the
asymmetric parameter $\alpha$ is tuned from $0$ to $\infty$. From
now on, we fix $J_l$ and $J_r$ to be unity, and focus on the effect
of the asymmetry of the rung interaction.

Previous studies for this model based on the DMRG simulations leads
to somewhat controversial results: Nishimoto and Arikawa suggested
that the spin gap vanishes as soon as the infinitesimally small
asymmetry is introduced~\cite{Nishimoto}, while Sakai {\it et.
el.}~\cite{Sakai} argued that a finite spin gap appears in a narrow
region around the rung-symmetry line and the phase transitions
between the gapless and gapful phases belong to the KT universality
class.

\section{numerical simulation}

For a quantum system described by the Hamiltonian $H(\alpha)$, with
$\alpha$ being the control parameter, the fidelity $F(\alpha_1,
\alpha_2) \equiv |\langle \Psi (\alpha_2) |\Psi (\alpha_1) \rangle |
$ between two ground states $|\Psi (\alpha_1) \rangle$ and $|\Psi
(\alpha_2) \rangle$ scales as $F(\alpha_1, \alpha_2) \sim
d(\alpha_1, \alpha_2)^L$. Here,  $d(\alpha_1, \alpha_2)$ is the
ground-state fidelity per lattice site, which characterizes how fast
changes when the thermodynamic limit is approached~\cite{zhou,zov}.
It satisfies the properties inherited from the fidelity $F(\alpha_1,
\alpha_2)$: (i) normalization $d(\alpha, \alpha)=1$; (ii) symmetry
$d(\alpha_1, \alpha_2)=d(\alpha_2, \alpha_1)$; and (iii) range $0
\leq d(\alpha_1, \alpha_2)\leq 1$. The ground-state fidelity per
lattice site exhibits singularities as a pinch point when the
control parameter $\alpha$ crosses a transition point $\alpha_c$.

%%%%%%%%%%%%%%%%%%%%%%%%%Fig4%%%%%%%%%%%%%%%%%%%%%%%%%%%%%%%%%%%
\begin{figure}
\centering
\includegraphics[width=3in]{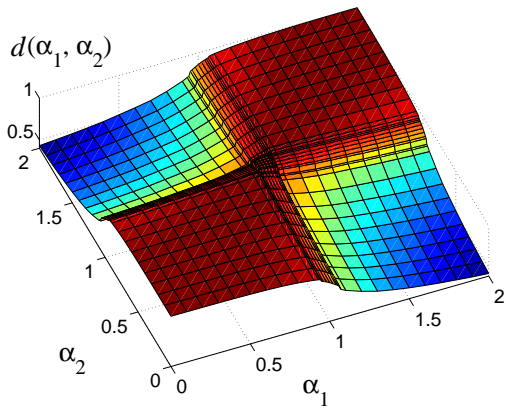}
\caption{(Color online) The ground-state fidelity per lattice site,
$d(\alpha_{1},\alpha_{2})$, for the spin 1/2
 three-leg Heisenberg tube with truncation dimension $\chi=6$. The
fidelity surface clearly indicates that there are two transition
points, as two pinch points occur at $\alpha_{c1} \sim 0.95$ and
$\alpha_{c2} \sim 1.06$, respectively.}\label{Fidelity3D}
\end{figure}
%%%%%%%%%%%%%%%%%%%%%%%%%%%%%%%%%%%%%%%%%%%%%%%%%%%%%%%%%%%%%%%%%%

The TN algorithm for quantum spin tubes, as described, allows to
efficiently compute the ground-state fidelity per lattice site. In
Fig.~\ref{Fidelity3D}, the ground-state fidelity per lattice site,
$d(\alpha_{1},\alpha_{2})$, is plotted for the spin 1/2 three-leg
Heisenberg tube, with truncation dimension $\chi=6$. Two transition
points are identified, as two pinch points on the fidelity surface
clearly indicates that there are two phase transitions at
$\alpha_{c1} \sim 0.95$ and $\alpha_{c2} \sim 1.06$, respectively.
Between two phase transition points, a finite gap opens, which
follows from the Lieb-Schultz-Mattis theorem, due to spontaneous
symmetry breaking of the translational symmetry, as reflected in the
two-site reduced density matrices. Given that both gapless phases
are Tomonaga-Luttinger liquids (see below for a detailed
discussion), the phase transitions between the gapless and gapful
phases belong to the KT universality class. Our results support the
DMRG results by Sakai {\it et. el.}~\cite{Sakai}.

\begin{figure}[!htp]
    %\centering
\includegraphics[width=3in]{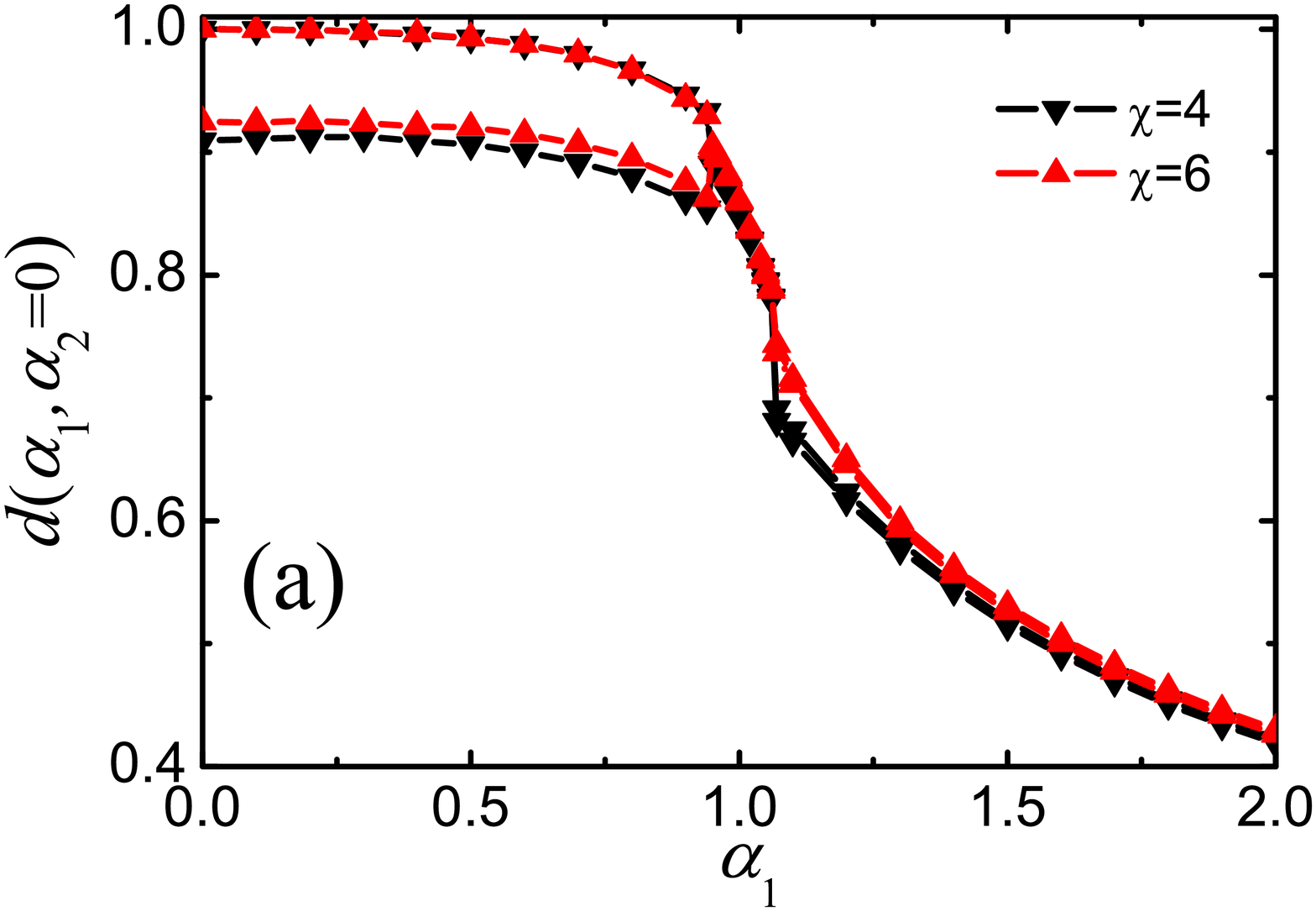}
\includegraphics[width=3in]{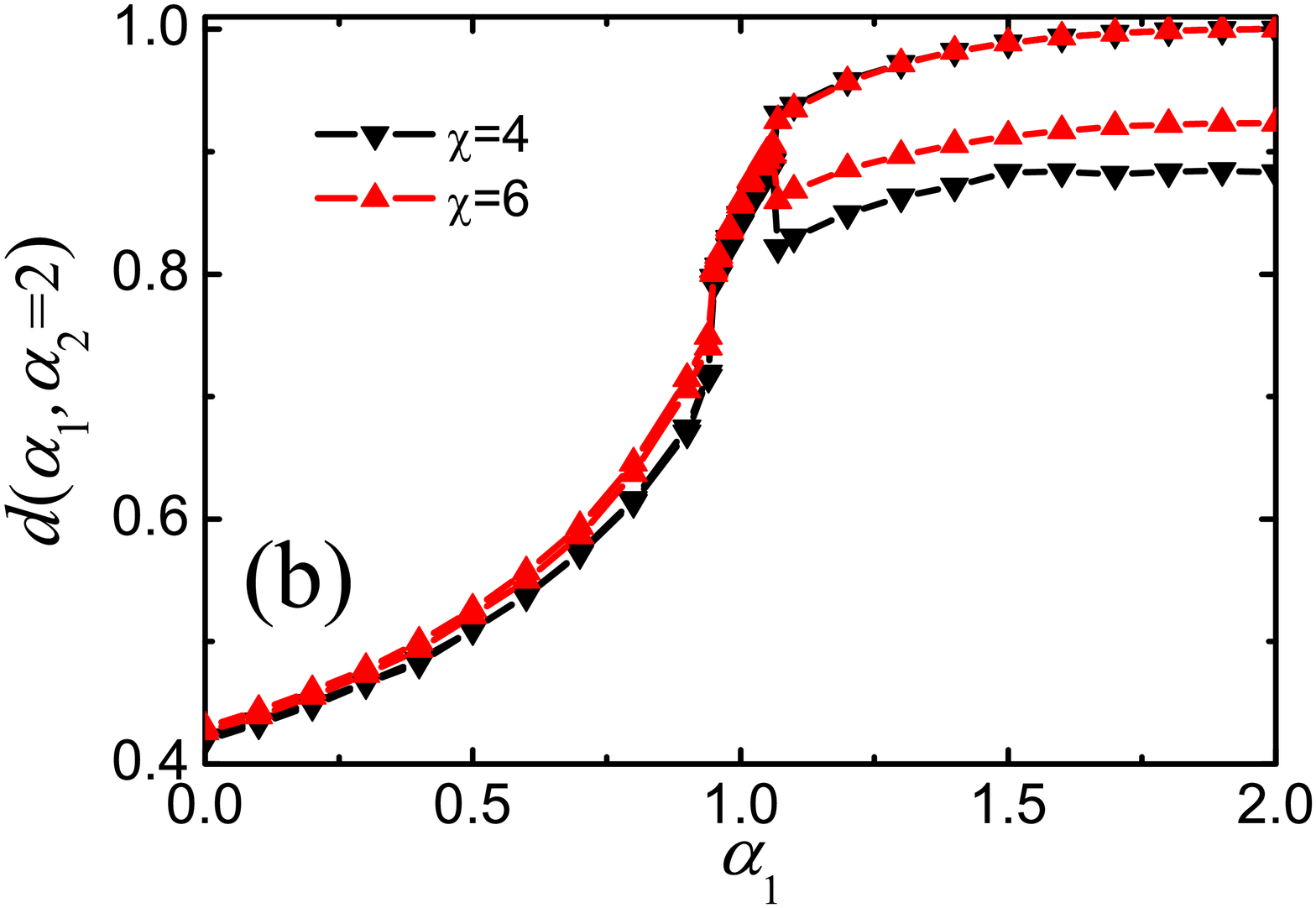}
\caption{(Color online) The ground-state fidelity per lattice site,
$d(\alpha_{1},\alpha_{2})$, as a function of $\alpha_1$ for a fixed
$\alpha_2$.  The truncation dimension is fixed as $\chi=4$ and
$\chi=6$, respectively.  Here, the reference state has been chosen:
(a) $\Psi(\alpha_{2}=0)$ and (b)
$\Psi(\alpha_{2}=2)$.}\label{Fidelity2D}
\end{figure}

In Fig.~\ref{Fidelity2D}, we plot the ground-state fidelity per
lattice site, $d(\alpha_{1},\alpha_{2})$, as a function of
$\alpha_1$, for a fixed $\alpha_2$, with different truncation
dimensions $\chi$. Here, two reference states $\Psi(\alpha_{2}=0)$
and $\Psi(\alpha_{2}=2)$ in the pseudo-$SU(2)$-symmetry-broken
phases have been chosen, respectively~\cite{wang}. The simulation is
performed for a randomly chosen initial state. The algorithm
automatically produces degenerate ground states which pseudo-break
the $SU(2)$ symmetry in the Tomonaga-Luttinger liquid phases. In the
pseudo-symmetry-broken phase, the ground-state fidelity per lattice
site, $d(\alpha_{1},\alpha_{2})$, exhibits different values lying
between two (extreme) degenerate ground states. In the symmetric
phase, $d(\alpha_{1},\alpha_{2})$ yields just one value. Thus, a
bifurcation point occurs, which coincides with the transition point
$\alpha_c$. Therefore, two phase transition points at $\alpha_{c1}
\sim 0.95$ and $\alpha_{c2} \sim 1.06$ are identified from the
bifurcation in the ground-state fidelity per lattice site. A
smoking-gun signature of the Tomonaga-Luttinger liquids is that such
a bifurcation tends to disappear when the truncation dimension
$\chi$ goes to infinity~\cite{wang}, as required to keep consistent
with the Mermin-Wagner theorem~\cite{mw} that no continuous symmetry
is spontaneously broken in a one-dimensional quantum system. This
implies that the degenerate ground states arise from the finiteness
of the truncation dimension, an artifact of the algorithm. However,
the singularity associated with the bifurcation still remains (as an
essential singularity as it should be for the KT transitions), even
if the truncation dimension goes to infinity. Notice that the
Goldstone modes survive as gapless excitations in the limiting case
without truncation.
%%%%%%%%%%%%%%%%%%%%%%%%%Fig5%%%%%%%%%%%%%%%%%%%%%%%%%%%%%%%%%%%
%%%%%%%%%%%%%%%%%%%%%%%%%%%%%%%%%%%%%%%%%%%%%%%%%%%%%%%%%%%%%%%%%%
\section{conclusion}
In this paper, an efficient algorithm has been developed for quantum
spin tubes in the context of the TN representations. It allows to
efficiently compute the fidelity per lattice site, which enables us
to identify quantum criticalities for quantum spin tubes. Our
simulation results for the isosceles spin 1/2 antiferromagnetic
three-leg Heisenberg tube suggest that two KT transitions occur as
the degree of asymmetry of the rung interaction is tuned, thus
offering an alternative route towards a resolution to the
conflicting results on this issue arising from the DMRG simulations.

\section*{ Acknowledgements} The work is supported by the National
Natural Science Foundation of China (Grant No: 10874252) and the
Fundamental Research Funds for Central Universities (Project No.
CDJXS11102213).

 %%%%%%%%%%%%%%%%%%%%%%%%%%%%%%%%%%%%%%%%%%%%%%%%%%%%%%%%%%%%

\end{document}